\documentclass[letter]{aa}  

\usepackage{graphicx}
\usepackage{txfonts}
\usepackage{placeins}
\usepackage{hyperref}
\hypersetup{
    colorlinks   = true,
    urlcolor     = cyan,
    linkcolor    = blue,
    citecolor   = blue
}
\usepackage{amssymb,amsfonts}
\usepackage{mathrsfs}%
\usepackage{xcolor}%
\usepackage{textcomp}%
\usepackage{manyfoot}%
\usepackage{booktabs}%
\usepackage{algorithm}%
\usepackage{algorithmicx}%
\usepackage{algpseudocode}%
\usepackage{listings}%

\newcommand{\msun}{M$_{\odot}$}
\newcommand{\TTunits}{Gyr$^{-2}$}
\newcommand{\reff}{R_{\rm eff}}
\newcommand{\leff}{\lambda_{\rm eff}}
\newcommand{\jwst}{JWST}
\newcommand{\bhfrac}{N_{\rm BH}/N}
\newcommand{\fboff}{\texttt{fbOFF}}
\newcommand{\fbon}{\texttt{fbON}}

\begin{document}

\title{The evolution of high-$z$ proto-star clusters into local globular clusters}

\author{
    A. Della Croce\inst{1}\corrauth{adellacr@iu.edu} \and
    E. Vesperini\inst{1}\email{evesperi@iu.edu} \and 
    R. Pascale\inst{2}\email{raffaele.pascale@inaf.it} \and
    A. Askar\inst{3}\email{askar@camk.edu.pl} \and
    F. Calura\inst{2}\email{francesco.calura@inaf.it} \and
    E. Dalessandro\inst{2}\email{emanuele.dalessandro@inaf.it} \and
    M. Giersz\inst{3}\email{mig@camk.edu.pl}
}
\institute{
    Department of Astronomy, Indiana University, Swain West 727 E. 3rd Street, Bloomington, 47405, IN, USA \and
    INAF, Osservatorio di Astrofisica e Scienza dello Spazio di Bologna, Via Gobetti 93/3, Bologna, 40129, Italy \and
    Nicolaus Copernicus Astronomical Center, Polish Academy of Sciences, ul. Bartycka 18, Warsaw, 00--716, Poland
}

\date{Received 24 April 2026 / Accepted 24 June 2026}

\abstract{
The James Webb Space Telescope (\jwst) detected numerous massive and relatively compact stellar clumps around proto-galaxies at high redshift ($z>0.5$). 
Their properties suggest that these systems may represent proto-globular clusters (GCs), but 
their possible connection to local old GCs is poorly understood.
In this Letter, we explore the dynamical evolution of proto-star clusters, building the missing evolutionary link between high-$z$ systems observed by JWST and local GCs. Our simulations include the effects of stellar interactions, stellar evolution, and the strong time-dependent cosmological tidal field in which these proto-star clusters evolve. We also explore the role of multiple stellar populations and stellar-mass black holes (BHs), two fundamental ingredients in stellar cluster dynamics.
We show that systems hosting multiple populations (as routinely observed in local GCs) are more likely to endure the early strong tidal field than single-population clusters. In addition, after 12 Gyr, such systems have properties consistent with those of Galactic GCs. Our work confirms that the high-$z$ clumps observed by \jwst~can be the progenitors of the local GCs. 
Finally, we show that a population of stellar-mass BHs within a proto-star cluster favors its disruption, but that surviving systems can retain a sizable population of BHs.
}

\maketitle
\nolinenumbers

\section{Introduction}
In recent years, the James Webb Space Telescope (\jwst)
has enabled the discovery of relatively massive ($10^{5}-10^7~$\msun) and dense ($10^3-10^5~$\msun pc$^{-2}$) stellar structures (referred to as clumps) at high redshift ($z\gtrsim 0.5$) thanks to the magnification of gravitational lensing \citep{mowla_etal2024,vanzella_etal2023,Claeyssens_etal2023,adamo_etal2024,messa_etal2026,whitaker_etal2026,abdurrouf_etal25}.
The spectral energy distributions of these clumps suggest that a fraction of them formed at very high redshift $z\simeq 8-12$ (about $12-13$ billion years ago) when the host galaxy was still assembling \citep{Claeyssens_etal2023,adamo_etal2024}.
The combination of these properties led to the tantalizing interpretation that such clumps might be the progenitors of globular clusters (GCs) in the local Universe.
Studying the physical nature of the "clumps" discovered by JWST and their possible link to local GCs would therefore lay the foundation for the theoretical framework to understand the origin of GCs and their role in galaxy assembly and evolution.

The detailed interplay among the numerous physical processes involved in the formation and evolution of star clusters remains an area of intense scientific investigation, both from theoretical \citep[e.g.,][for a review]{krause_etal2020} and observational \citep[e.g.,][for a review]{adamo_etal2020} perspectives. However, most studies so far have focused on the local Universe, where star clusters can be resolved into individual stars.
While these systems provide valuable insights, the possibility of studying "in real time" the formation of the most massive clusters and their natal environment is unprecedented.

\jwst ~observations thus sparked a novel interest in the formation of star clusters within large-scale cosmological simulations, pushing the spatial and mass resolutions to resolve their emergence \citep{kimm_etal2016,ma_etal2020,calura_etal2022,calura_etal2025,reinacampos_etal2022,sameie_etal2023,garcia_etal2023,pascale_etal2025,taylor_etal2025_nature}.
However, due to their high computational cost, zoom-in cosmological simulations are typically performed only down to redshifts $z\sim 6-5$. As a result, they cannot address the fundamental question of whether the high-redshift clumps observed by JWST are the actual progenitors of present-day GCs. 

Recent studies showed that most clusters would undergo rapid disruption in the strong cosmological tidal fields \citep{li_gnedin2019,webb_etal2024}.
Surviving systems lose a considerable fraction of their initial mass, attaining structural properties different from those of the majority of Galactic GCs \citep{rodriguez_etal2023,giunchi_etal2025}.
In this Letter, we build upon previous studies to address the dynamical evolution of clumps from their birth times ($z\gtrsim8$, roughly 12~Gyr ago) down to the present day.
Within this context, previous works either included semi-analytic prescriptions for the cluster dynamical evolution \citep{li_gnedin2019,meng_gnedin2022} or focused on the cluster population of a particular galaxy in a specific cosmological simulation \citep{rodriguez_etal2023}. The scope of the present work is complementary: to present a possible evolutionary pathway of the high-$z$ clumps observed by JWST into local GCs.

Our investigation is based on a survey of Monte Carlo simulations following the structural and dynamical evolution of star clusters. We ran the simulations with MOCCA \citep{giersz_etal2013, hypki_etal2013, hypki_etal2022}, a Monte  Carlo code that accurately treats the evolution of single and binary stars, the effects of stellar encounters,  and the presence of an external tidal field. In this respect, we modified MOCCA to handle a time-dependent tidal field \citep[as found in cosmological simulations, e.g.,][]{kruijssen_etal2015,li_gnedin2019,meng_gnedin2022}. 
Details on the cosmological tidal field adopted are discussed in Appendix~\ref{sec:MOCCAtt}.
We also studied the evolution of systems hosting either a single (SP) or multiple stellar populations (MPs, as routinely observed in local, old GCs, see e.g., \citealt{gratton_etal2019} and references therein).
In contrast to SP systems, MP systems have two populations with different initial spatial distributions.
This choice of initial structural properties is informed by numerous hydrodynamical simulations that predict second population (2P) stars form more centrally concentrated than first population \citep[1P,][]{dercole_etal2008,calura_etal2019,lacchin_etal2022}.  Imprints of such primordial spatial differences are observed in several Galactic GCs \citep{dalessandro_etal2019,dalessandro_etal2024,leitinger_etal2023}.
We refer the reader to Appendix~\ref{sec:numerical_methods} for further details on the simulation initial conditions. 

\section{Results}\label{sec:results}
\begin{figure*}[!th]
    \centering
    \includegraphics[width=0.4\linewidth]{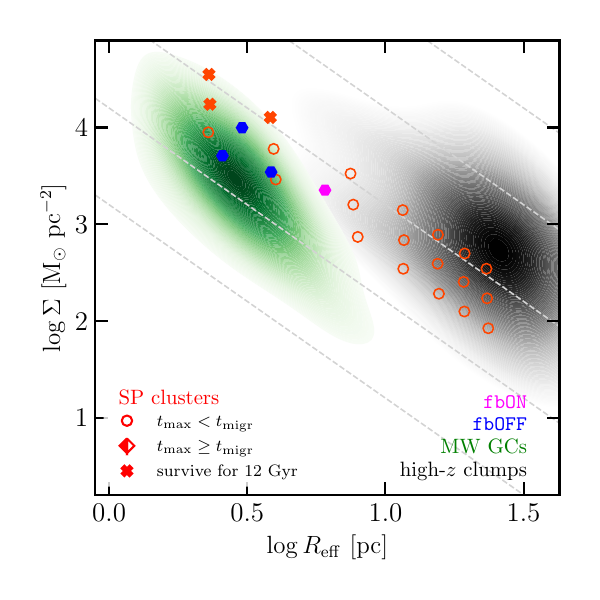}
    \includegraphics[width=0.4\linewidth]{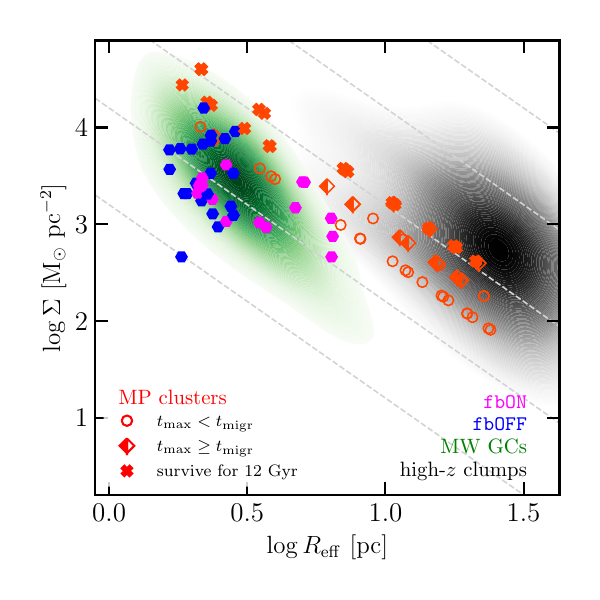}
    \caption{Surface density -- effective radius plane. 
    The left panel shows SP models, while the right one presents systems with MPs. 
    In red are simulations' initial conditions: empty circles represent simulations that dissolved before migration (labeled as $t_{\rm max}<t_{\rm migr}$ with $t_{\rm max}$ the system survival time). Half-filled diamonds are simulations that survived the early phase but dissolved before the present day ($t_{\rm max}\geq t_{\rm migr}$).
    Finally, red crosses present the simulations that survive for 12 Gyr, with their final positions marked by hexagons: blue markers show simulations without the fallback prescription for BH formation (referred to as \texttt{fbOFF}), while simulations including it (\texttt{fbON}) are in magenta.
    Green contours show the location of MW GCs (data are from \citealt{baumgardt_etal2020} catalog), and black contours show the distribution of young ($\leq50$~Myr) clumps (latest compilation by \citealt{claeyssens_etal2026}) that trace the early Universe star formation conditions. 
    Dashed lines in gray represent iso-mass lines respectively $10^4, 10^5, 10^6, 10^7, 10^8$~\msun.
    }
    \label{fig:logsigma_reff_wEvolution12Gyr}
\end{figure*}

Massive star clusters form in dense and high-pressure regions, where the tidal field is stronger and changes rapidly with time due to gas accretion and encounters with giant molecular clouds, before large-scale potential variations (e.g., triggered by galaxy infalls and mergers) remove the cluster from the gaseous disk and inject it into the halo of the host \citep{kruijssen_etal2015,li_gnedin2019,meng_gnedin2022}. We thus first identify systems that survive long enough to escape their birthplace and those that will eventually survive for a Hubble time.
We considered a system to have survived the early phases of a stronger, highly time-dependent tidal field if it retained at least $10^{-2}$ of the initial cluster mass at the time of migrating from the strong tidal field regime (mimicking the cluster formation environment) to a weaker one (characteristic of halo-like orbits for MW GCs). 
The timing of the Gaia-Enceladus merger event guided the choice of the transition time between the two regimes \citep{helmi_etal2018,belokurov_etal2018}.   
The same definition was adopted at 12 Gyr to identify systems that survived to the present day.

\textit{Clusters with a single stellar population --} 
We found that only the most compact ($r_{\rm h}\leq 5$~pc) and most massive ($N \geq 10^6$) SP clusters considered in this work retain enough mass to survive the initial phase of a stronger tidal field.
This is shown in the left panel of Figure~\ref{fig:logsigma_reff_wEvolution12Gyr}, where initial conditions are compared with high-$z$ clump observations \citep{claeyssens_etal2026} and their properties after 12~Gyr with the Milky Way (MW) GCs population \citep{baumgardt_etal2020}. Interestingly, the handful of systems that survive evolve towards the bulk of MW GCs in the $\log\Sigma - \reff$ plane\footnote{Throughout this work, we defined $\Sigma \equiv M_{\rm cl}/2\pi\reff^2$, with $M_{\rm cl}$ and $\reff$ being the total clump/cluster mass and half-light radius, respectively. The same definition was adopted for all the stellar systems discussed in this work (being them high-$z$ clumps, MW GCs, or numerical simulations).}.
However, they are not really representative of the observed clump population (see Figure~\ref{fig:logsigma_reff_wEvolution12Gyr}), which extends to much larger sizes ($\reff \gtrsim 10$ pc). 
SP proto-star clusters starting with values of $\Sigma$ and $\reff$ in the range of those observed by JWST do not survive the strong tidal fields at early times.

\textit{Clusters hosting multiple populations --}
JWST observations allow us to derive masses ($M_{\rm cl}$) and half-light radii ($\reff$) of high-$z$ lensed clumps. Although these values are useful quantities, they do not provide detailed information on the clump density profiles or internal substructure, for example, resulting from multiple star-formation episodes. The details of the clusters' initial structural properties are critical to determine their early survival and dynamical evolution. In particular, as we discuss below, the more complex multi-scale structure of MP GCs favors the survival of a GC with the same mass and size, even though a large fraction of the first-population (1P) stars (with pristine chemical composition) are lost in the early stages of evolution.
The set of simulations with MPs has the same observable properties ($M_{\rm cl},\reff$) as the SP ones, but is characterized by a denser, more centrally concentrated subsystem of second-population (2P) stars (representing the chemically enriched population) embedded in a more extended 1P cluster. 

The right panel of Figure~\ref{fig:logsigma_reff_wEvolution12Gyr} shows the $\log\Sigma - \reff$ plane for the suite of MPs simulations.
The first remarkable result is that many more systems survive, including those with $\reff > 10$ pc, which nicely overlap with the observed high-$z$ clump population. 
The second result is that after 12 Gyr of stellar and dynamical evolution, the clumps match the values of $\Sigma$ and $\reff$ (and therefore of mass) of the MW GC population. 
This suggests that the high-$z$ clumps observed by \jwst~may indeed evolve into the old and massive star clusters we observe in the local Universe if they host MPs. This highlights the central role of MPs: systems hosting MPs are more likely to survive the strong tidal fields during the early evolutionary phases when the host galaxy is still building up. 

\begin{figure*}[!th]
    \centering
    \includegraphics[width=0.4\linewidth]{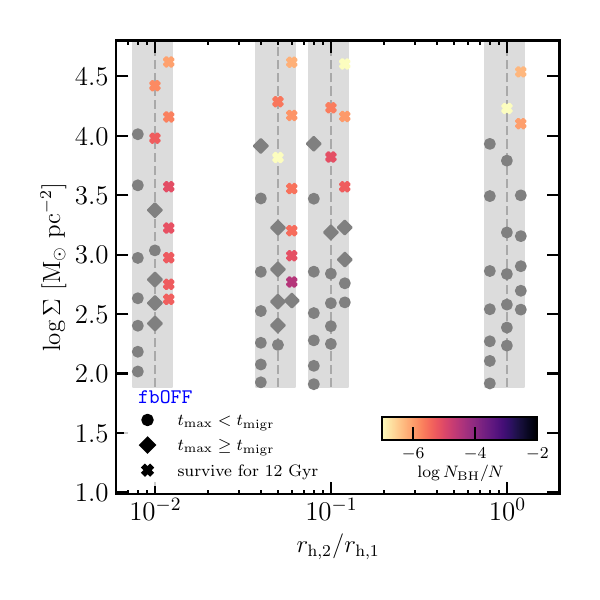}
    \includegraphics[width=0.4\linewidth]{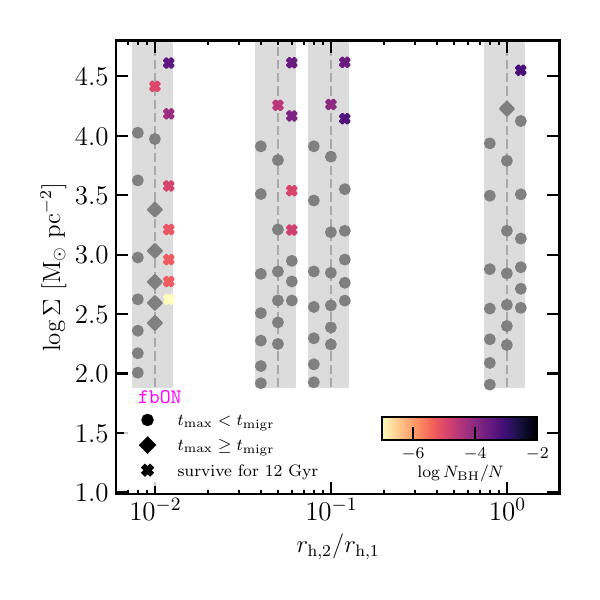}
    \caption{Mass-surface density -- concentration plane presenting the simulation initial conditions. 
    Simulations with different prescriptions for BH formation are presented separately in different panels: on the left are \texttt{fbOFF} simulations, while \texttt{fbON} simulations are in the right panel.
    The vertical gray bands enclose simulations with the same concentration but different numbers of particles (SP systems have $r_{\rm h,2}/r_{\rm h,1}=1$). From left to right, $5\times10^5$, $10^6$, and $2\times10^6$. 
    In addition, simulations with the same $r_{\rm h,2}/r_{\rm h,1}$ and mass (i.e., vertical sequences) are naturally ordered by increasing initial half-mass radius from top to bottom.
    We present systems that survive different stages with different markers, while proto-GCs that survive for 12~Gyr are color-coded according to the BH number fraction at 12~Gyr. Other simulations are shown in gray.
    }
    \label{fig:logsigma_concentration_BHs}
\end{figure*}

This final point is further highlighted in Figure~\ref{fig:logsigma_concentration_BHs}, where the survival of MP systems with different relative concentrations and SP ones is directly compared.
First, progressively moving to systems with higher 2P concentration (i.e., lower $r_{\rm h,2}/r_{\rm h,1}$) while keeping the same values for all the other cluster parameters (e.g., mass, radii, and tidal field) favors the survival of the system. While a handful of SP clusters (i.e., $r_{\rm h,2} = r_{\rm h,1}$) survive, systems hosting MPs are more likely to survive, even those with large extensions (e.g., $\reff > 10$ pc) for sufficiently high concentrations (e.g., $r_{\rm h,2}/r_{\rm h,1} = 0.01$). 
Some of these systems are initially overfilling their Jacobi surface (within the adopted tidal field strength, see Appendix~\ref{sec:MOCCAtt}) and thus lose many stars in the initial stages. 
However, escaping stars are mainly 1P stars, while centrally concentrated 2P stars are preferentially retained. In general, it is this early episode of loss of 1P stars that increases the 2P-to-1P ratio in clusters \citep{vesperini_etal2021} to values broadly consistent with the range of present-day fractions observed in Galactic GCs \citep[e.g.,][]{milone_etal2017}.

\textit{The role of stellar-mass BHs --}
The presence of stellar-mass BHs within a star cluster has strong implications for the system's dynamical evolution \citep{mackey_etal2007,breen_heggie2013,arcasedda_etal2018,giersz_etal2019,kremer_etal2020,dellacroce_etal2024}. At the same time, massive star clusters are primary sources of gravitational waves as repeated interactions between a binary system of compact objects and single objects harden the binary, accelerating its in-spiral and favoring the merger \citep{rodriguez_etal2016_BBH,hong_etal2018}. 
Addressing the role of BHs in the early cluster evolution and exploring the BH populations for systems that survive for a Hubble time are thus timely science cases. 

Different panels in Figure~\ref{fig:logsigma_concentration_BHs} present systems with different prescriptions for the formation of BHs. 
In the simulations presented in the left panel, the newly formed BH does not re-accrete material expelled in the supernova explosion (hereafter labeled as \fboff), implying high natal kicks (adopted to be the same as neutron stars, \citealt{hobbs_etal2005}). On the other hand, simulations shown in the right panel include the prescriptions for fallback material by \citet[][\fbon]{Belczynski_etal2002}, resulting in significantly more BHs retained thanks to lower BH natal kicks.
We considered both possibilities to explore the role of BH retention in the early evolution and survival of proto-GCs.
In this respect, a clear difference emerges in Figure~\ref{fig:logsigma_concentration_BHs}: the presence of a large number of BHs within the system favors their disruption.
This is due to the heating of the BH population on the stellar component, which, over time, prevents the system from contracting and exposes the cluster to the tidal field, thereby accelerating its dissolution \citep{giersz_etal2019,kremer_etal2020}.

The properties of the BH population at 12~Gyr are presented in Figure~\ref{fig:logsigma_concentration_BHs}, where each surviving cluster is color-coded according to the BH number fraction at 12~Gyr ($\bhfrac$). This is defined as the ratio of the number of BHs ($N_{\rm BH}$) to the total number of bound stars ($N$), quantifying the influence of BHs on the stellar component.
Focusing on \fbon~systems (the right panel in Figure~\ref{fig:logsigma_concentration_BHs}), initially compact clusters generally attain higher $\bhfrac$ than extended ones. These systems indeed have room to expand within their Jacobi surface under the influence of the BH population, and their half-mass relaxation time quickly surpasses the cluster age, halting BH ejections mediated by dynamical interactions. On the other hand, more compact and denser systems (e.g., $\reff\lesssim1~$pc) would retain fewer BHs due to more efficient removal by dynamical interactions.
Extended systems ($\reff\gtrsim 10~$pc, see Figure~\ref{fig:logsigma_concentration_BHs}) are carved from the tidal field, as 1P stars initially extend beyond the Jacobi radius.
Mass loss due to stellar evolution and the loss of stars caused by the cluster's expansion further limits the system extension as the tidal radius shrinks.
Both processes lower the half-mass relaxation timescale, which regulates the BH ejection rate \citep{kremer_etal2019}.

Opposite to \fbon~systems, differences in the BH fraction among \fboff~clusters are driven by the overall number of bound stars rather than the number of BHs. Therefore, initially more compact clusters end up with higher $N$, and thus lower $\bhfrac$ than extended clusters (although a few more BHs could be retained by more compact clusters thanks to a deeper potential well).

\section{Conclusions}
Starting from a large grid of numerical simulations that account for internal dynamics and stellar evolution evolving within cosmological tidal fields, we 
presented the survival and dynamical evolution of typical high-$z$ proto-star clusters.
The main result of this work is that clumps observed at high redshift can indeed evolve into GC-like objects.
However, it emerges that
\begin{enumerate}
    \item The key ingredient to endure the strong tidal fields of the cluster natal environment is an initial structure characterized by the presence of a centrally concentrated stellar component. Such a structure is linked to the presence of multiple stellar populations 
    and motivated by theoretical studies of MP star clusters that predict the 2P forms in a more centrally concentrated subsystem within the more extended 1P system;
    \item SP clusters without internal substructure are more likely to dissolve during their early evolution and might survive only if they were extremely compact and massive initially;
    \item Finally, systems that form early in the Universe can retain significant fractions of BHs for their entire existence up to the present day.
\end{enumerate}
This work is the first in a series in which we quantitatively explore the possible link between the high-z clumps observed by
JWST and the local GC population. In future works, we will focus on the clusters' internal dynamical properties and their stellar content.

\begin{acknowledgements}
We thank the referee for their comments that helped improve the clarity of the paper.
We thank A. Claeyssens for privately sharing the data used in Figure~\ref{fig:logsigma_reff_wEvolution12Gyr}.
R.P. acknowledges financial support from the INAF Minigrant programme (OF 1.05.24.07.05, CUP C33C24001390005) within the framework of Astrofisica Fondamentale of the Istituto Nazionale di Astrofisica (INAF).
AA acknowledges that this research was funded in part by the National Science Centre (NCN), Poland, grant No. 2024/55/D/ST9/02585. For the purpose of Open Access, the author has applied a CC BY public copyright license to any Author Accepted Manuscript (AAM) version arising from this submission.
E.D. acknowledges financial support from the INAF Data analysis Research Grant (PI E. Dalessandro) of the “Bando Astrofisica Fondamentale 2024”.
MG was supported by the Polish National Science Center (NCN) through the grant 2021/41/B/ST9/01191.
\end{acknowledgements}

\bibliographystyle{aa}
\bibliography{bibliography}

\onecolumn
\appendix
\section{Numerical simulations initial conditions}\label{sec:numerical_methods}
For the initial conditions of our simulations, we considered systems with $N = [0.5,1,2]\times 10^6$ particles and half-mass radii $r_{\rm h} = 3,5,10,15,20,25,30$ pc. The adopted initial conditions allow us to model very-massive clusters ranging from small to large extensions that match the variety of high-$z$ clump properties probed by \jwst~\citep{mowla_etal2022,mowla_etal2024,vanzella_etal2023,Claeyssens_etal2023,adamo_etal2024,messa_etal2026,abdurrouf_etal25}. 
This is particularly relevant as it opens the possibility of investigating the dynamical evolution of candidate proto-GCs, assessing their survival in a cosmological environment. Nonetheless, we note here that observed clumps can attain much larger sizes ($\gtrsim 100$ pc) and masses ($\sim 10^7$~\msun). Such extended clumps, however, are most likely not individual star clusters, but rather star-forming complexes that may host a compact stellar cluster embedded (or possibly even more than one). Concerning the masses, while larger clusters are certainly possible, simulating a collisional system with about $10^7$ particles is computationally challenging, especially when exploring multiple initial conditions.

All systems were initialized as King \citep{king_1966} models with $W_0 = 5$ and masses were sampled from a Kroupa \citep{kroupa_2001} initial mass function (with stellar masses ranging from 0.08\msun~to 150\msun) with metallicity $Z=0.001$.
This choice, combined with the total number of particles, produces systems with masses $2.93\times10^5$ \msun, $5.86\times10^5$ \msun, and $1.17\times10^6$ \msun.
Additionally, for each setup, two different simulations were run, either implementing (\fbon) or not (\fboff) the fallback prescription for BH formation \citep{Belczynski_etal2002}.
Within this framework, a fraction of the supernova ejecta falls back onto the newly formed BH, thereby reducing its natal kick through linear momentum conservation. Therefore, \fbon~simulations retain a higher fraction of the BHs formed than the \fboff~case, allowing us to explore the role of early BH retention in the system survival.

In addition to the aforementioned initial conditions, we considered the possibility that proto-star clusters host MPs. 
The motivation behind this approach is dual. First, most (if not all) GCs we observe in the local Universe host MPs \citep{piotto_etal2015,milone_etal2017,bastian_lardo_2018,gratton_etal2019}.
Hence, if clumps observed at high-$z$ are actually the progenitors of local GCs, they likely host MPs as well. 
On the observational side, \jwst~provides integrated properties of the clumps (such as mass and effective radius), but can not resolve any internal substructure. 
Building on these points, we also considered systems that host MPs.
In particular, following the results of several studies based on hydrodynamical simulations of MP formation \citep[e.g.,][]{dercole_etal2008,bekki_etal2010,calura_etal2019,lacchin_etal2022}, we adopt initial conditions characterized by a 2P subsystem centrally concentrated in the inner regions of an extended 1P system. We note that some memory of these differences between the 1P and 2P spatial distributions has been found in many observational studies.
Still, additional theoretical and observational investigations are necessary to fully explore the range of possible initial 1P and 2P dynamical properties and their link with those observed in local GCs.

Using the same prescriptions as \citet{hypki_etal2022}, we constructed equilibrium systems composed of two populations with different relative masses and sizes. The overall system, however, had the same total mass and size 
as the corresponding SP model whose initial conditions were inspired by JWST observations. 
We considered three different values of relative concentrations, defined as the second-to-first population half-mass radii ratio $r_{\rm h,2}/r_{\rm h,1} = 0.1, 0.05, 0.01$. 
In all cases, $25\%$ of the initial GC mass is in 2P stars.
For any given number of particles (i.e., system's mass), fraction of 2P stars and relative concentration between the two populations, $r_{\rm h,1}$ (and therefore $r_{\rm h,2}$) is derived numerically so that the half-mass radius ($r_{\rm h}$) of the system is the same as that of the corresponding SP system. This naturally reflects in a hierarchy of scales: $r_{\rm h,2}<r_{\rm h}<r_{\rm h,1}$. Similarly to the half-mass, the effective radius (see e.g., Fig.~\ref{fig:logsigma_reff_wEvolution12Gyr}) is defined as the radius enclosing half of the total system light, regardless of the stellar population.
Besides the half-mass radius and the overall mass, there are no other additional differences between 1P and 2P stars: they are sampled from the same stellar initial-mass function and had an equal $W_0 (=5)$ King parameter. 
This choice was made to minimize the number of free parameters in the simulations, to better highlight the role of internal substructures.
Ultimately, the adopted initial conditions allow us to explore the survival and evolution of candidate proto-GCs observed at high-$z$, and also to investigate, for any observed pair of mass and size, the possible different evolutions due to the presence of BHs and primordial substructures in clusters hosting multiple stellar populations.

\section{Time dependent tidal fields in MOCCA}\label{sec:MOCCAtt}
\begin{figure}[!th]
    \centering
    \includegraphics[width=0.85\linewidth]{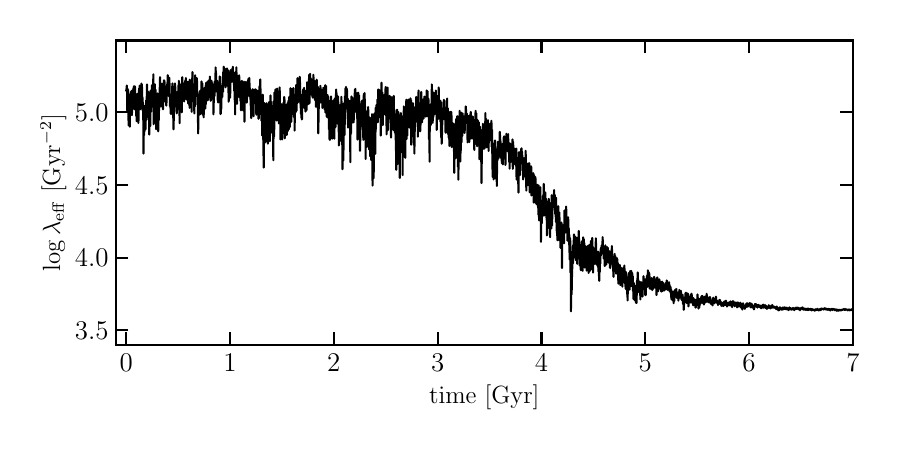}
    \caption{Time evolution of the mock tidal field strength ($\leff$) adopted in this work. $\leff$ directly enters in the definition of the Jacobi radius through eq.~\ref{eq:jacobi_radius}}
    \label{fig:lambda_simulationsurvey}
\end{figure}

We define the strength of the tidal field through the formalism introduced by \citet{renaud_etal2011}. In particular, we use the effective eigenvalue ($\leff$) of the tidal tensor \citep{pfeffer_etal2018} to define the Jacobi radius of a spherical system as a function of time
\begin{equation}
    r_{\rm J}(t) \equiv \left( \frac{G M_{\rm cl}(t)}{\leff(t)}\right)^{1/3}\,.
    \label{eq:jacobi_radius}
\end{equation}
The time evolution of tidal field strength could be derived directly from cosmological simulations of galaxy formation \citep{renaud_etal2011,li_gnedin2019,rodriguez_etal2023}. However, this approach is limited to the specific star formation history selected and cannot provide us with a broader (in a statistical sense) view of the survival of star clusters evolving in the same environment. Building upon the work by \citet{webb_etal2024}, we generated a realistic and time-dependent tidal field from time series. In particular, we adopted the analytical functional form
\begin{equation}
    \leff(t) = A f(t) (1+\overline{\lambda}(t)) + B(R_{\rm G,eff})\,,
    \label{eq:lambda}
\end{equation}
where
\begin{enumerate}
    \item $A$ sets the initial amplitude/strength of the tidal field (in \TTunits);
    \item $\overline{\lambda}(t)$ is the inverse Fourier transform of $\tilde{\lambda}(\nu)$ whose power spectrum is $|\tilde{\lambda}(\nu)|^2 \propto \nu^{-\alpha}$;
    \item $B \equiv 3 G M_{\rm MW}(<R_{\rm G,eff})/R_{\rm G,eff}^3$ (under the assumption of a circular orbit at the effective Galactocentric radius $R_{\rm G,eff}$, \citealt{baumgardt_makino2003,cai_etal2016});
    \item and $f(t)$ is the additional temporal dependence such that $f(0)=1$ (i.e., $\leff(0) = A +B$) and $f(t\to\infty) = 0$ (i.e., $\leff(t\to\infty) = B$)\,.
\end{enumerate}
We adopted
\begin{equation}
    f(t) = \frac{1}{1+(t/t_{\rm migr})^\chi} \quad {\rm with}~\chi>0\,,
\end{equation}
which allows for an initial stronger tidal field.

Equation~\ref{eq:lambda} thus allows us to model a tidal field of arbitrary initial strength (regulated by $A$) that weakens over time to match the tidal field strength experienced by present-day GCs in the MW (we adopted the MW cumulative mass profile derived by \citet{cautun_etal2020} to estimate $M_{\rm MW}(<R_{\rm G,eff})$). 
Additionally, the tidal field has sharp time variations introduced by $\overline{\lambda}(t)$ (with the frequency range regulated by its power spectrum) whose amplitude depends on the local (in time) amplitude of $f(t)$, mimicking the fast, temporal variation of the tidal field strength observed from cosmological simulations \citep{li_gnedin2019,rodriguez_etal2023}.

Finally, we adopted the following values
\begin{enumerate}
    \item $A=10^5$ \TTunits;
    \item $\alpha = 1$ following \citet{webb_etal2024};
    \item $R_{\rm G,eff} = 6$~Kpc (a representative value for the MW GCs, \citealt{baumgardt_makino2003,cai_etal2016}), corresponding to $\leff(t\to\infty) \simeq 4.26 \times 10^3$ \TTunits;
    \item $t_{\rm migr} = 3.5$~Gyr, and $\chi=10$ resulting in a steady tidal field for about $1.5$~Gyr and decreasing to half its initial value at $t_{\rm migr}$.
\end{enumerate}
In Figure~\ref{fig:lambda_simulationsurvey} we show the time evolution of $\leff$
obtained assuming the parameters above, which are common to all our simulations.
The mock tidal field adopted in this work has a longer phase of strong tidal forces than, for instance, the one reported by \citep{meng_gnedin2022}. However, if large-scale gravitational potential variations are responsible for removing proto-star clusters from the high-density regions of formation \citep{kruijssen_etal2015,li_gnedin2019}, the typical time scale would depend on the specific merger history of the galaxy.
The latest results about the MW assembly history suggest that our Galaxy experienced two significant mergers: the so-called Low-energy-Kraken-Heracles (about 1.5~Gyr after the Big Bang, \citealt{massari_etal2026}) and Gaia-Enceladus (about 10~Gyr ago, \citealt{helmi_etal2018,belokurov_etal2018,gallart_etal2019}). The adopted migration time results in a decrease of the tidal field strength roughly consistent with the time of this latter episode. 
Finally, all these assumptions together make the tidal field presented in Figure~\ref{fig:lambda_simulationsurvey} a realistic model for MW GCs.

It is essential to emphasize that accurately modeling stellar escape due to the tidal field in Monte Carlo codes is challenging (see, for example, the discussion in \citealt{sollima_mastrobuonobattisti2014,rodriguez_etal2023}), whereas direct-summation $N$-body codes can naturally account for external forces by explicitly including them in the equations of motion \citep{renaud_etal2011}.
The Monte Carlo method leverages our knowledge of star orbits in spherically symmetric clusters to speed up the computation, enabling surveys of simulations with a realistic number of stars.
However, the presence of a tidal field breaks the spherical symmetry. Therefore, Monte Carlo codes account for a tidal boundary (determined by $r_{\rm J}$) coupled with an escape criterion calibrated on direct-summation $N$-body simulations \citep{fukushige_heggie_2000}.
While modern codes (such as \texttt{MOCCA} and \texttt{CMC}) have been widely tested against direct $N$-body codes \citep{giersz_etal2013,rodgriguez_etal2016}, 
showing excellent agreement while remaining computationally efficient, their simplified description of external forces cannot capture the full complexity of stellar escape \citep{ross_etal1997,fukushige_heggie_2000,baumgardt_2001,madrid_etal2017}.

\end{document}